\documentclass[preprint,times,3p,twocolumn]{elsarticle}  
\usepackage{latexsym}
\usepackage{amssymb}
\usepackage{amsmath}
\usepackage{graphicx}
\usepackage{multirow}
\usepackage{xcolor}

\begin{document}
	
\begin{frontmatter}

\title{Evidence for maximality of strong interactions from LHC forward data}

\author[em]{E. Martynov}
\ead{martynov@bitp.kiev.ua}

\author[bn]{B. Nicolescu}
\ead{basarab.nicolescu@gmail.com}

\address[em]{Bogolyubov Institute for Theoretical Physics, Metrologichna 14b, Kiev, 03680 Ukraine}
\address[bn]{Faculty of European Studies, Babes-Bolyai University, Emmanuel de Martonne Street 1, 400090 Cluj-Napoca, Romania}




\begin{abstract}
It is important to check if the Froissaron-Maximal Odderon (FMO) approach is the only model in agreement with the LHC data. We therefore generalized the FMO approach by relaxing the $\ln^2s$ constraints both in the even-and odd-under-crossing amplitude. We show that, in spite of a considerable freedom of  a large class of amplitudes, the best fits bring us back to the maximality of strong interaction. Moreover, if we leave Odderon Regge pole intercept $\alpha_O(0)$ completely free we find a very good solution for $\alpha_O(0)$ near -1 in agreement with the result of oddballs spectroscopy in QCD based on AdS/CFT correspondence.
\end{abstract}


\end{frontmatter}

\section{Introduction}
In a previous paper \cite{MN}, we showed that the Froissaron-Maximal Odderon (FMO)  approach is in agreement with both the $\ln^2s$ behaviour of total cross sections at high energies and the surprisingly low TOTEM $\rho^{pp}$ datum at 13 TeV \cite{TOTEM}. 

The even-under-crossing amplitude in the FMO approach is constituted by a 2-component Pomeron (the Froissaron and the Pomeron Regge pole located at $j=1$) to which are added the secondary Regge poles located around $j=1/2$ of even signature. In its turn, the odd-under-crossing amplitude is constituted by a 2-component Odderon (the Maximal Odderon and the Odderon Regge pole located at $j=1$) to which are added secondary Regge poles located around $j=1/2$ of odd signature. The Odderon was introduced in 1973 on the theoretical basis of asymptotic theorems \cite{L-N}. It has to be noted that, when it was introduced, there were no experimental indications for its existence.

The Froissaron (or Maximal Pomeron) and the Maximal Odderon represent the asymptotic contributions (at $s\to \infty$), while the Regge poles describe the low and medium range of energies. In other words, the FMO approach describes the finite-energy effects of asymptotic theorems, by a nice and simple interplay between asymptotic and non-asymptotic contributions. 
 
The Froissaron corresponds to the maximal behavior $\ln^2s$ of total cross sections allowed by general principles. It was first introduced by Heisenberg in 1952 \cite{Heis} and a rigorous demonstration was given nine years later by Froissart \cite{Froissart}.

In its turn, the Maximal Odderon corresponds to the maximal behavior $\ln^2s$ of the real part of the odd-under-crossing amplitude and to the maximal behaviour $\ln s$ of the difference of the antihadron-hadron and hadron-hadron total cross sections \cite{L-N, Fischer}. 

The FMO approach embodies  a new form of the old principle of maximum strength of the strong interactions \cite{Chew}: both the even and the odd-under-crossing amplitudes saturate functionally the asymptotic bounds.

In the present paper we investigate the following question: is the maximality of strong interactions not only in agreement with experimental data but it is even required by them? For that we are relaxing the $\ln^2s$ constraint by a generalization of the FMO approach.

\section{Generalization of the FMO approach}

Let us consider the following form of the amplitudes:
 \begin{equation}\label{eq:Ft=0}
F^H_+(z)=i(s-2m^2)[H_1\ln^{\beta_F}(-iz)+H_2\ln^{\beta_F-1}(-iz)+H_3], 
\end{equation}
\begin{equation}\label{eq:MOt=0}
F^{MO}_-(z)=(s-2m^2)[O_1\ln^{\beta_{MO}}(-iz)+O_2\ln^{\beta_{MO}-1}(-iz)+O_3], 
\end{equation}
\begin{equation}\label{eq:sec Regge}
F^R_\pm(z)\equiv \binom{P,R_+}{O,R_-}=-\binom{1}{i}C^R_\pm (-iz)^{\alpha_\pm(0)}.
\end{equation}
where 
\begin{equation}\label{eq:z-def}
z=(s-2m^2)/2m^2
\end{equation}  
and $m$ is mass of proton.
The elastic $pp$ and $\bar pp$ forward scattering amplitudes are defined through the relations
\begin{equation}\label{eq;}
F_\pm (z,0) = (1/2) (F_{pp}(z,0) \pm F_{\bar pp}(z,0)). 			
\end{equation} 
The observables are 
\begin{equation}\label{eq:sigdef}
\sigma_{tot}(s) =\text{Im}F(z)/ \sqrt{(s (s-4m^2 )} 
\end{equation}
\begin{equation}\label{eq:rhodef}
\rho(s) = \text{Re}F(z)/\text{Im}F(z). 				
\end{equation}
The proton-proton forward scattering amplitude  is
\begin{equation}\label{eq:ppampl}
F_{pp}(z) = F^H_+(z) + F^{MO}_-(z) +F^P(z)+F^O(z)+ F^R_+(z) + F^R_-(z)
\end{equation}
and the amplitude of antiproton-proton scattering is			
\begin{equation}\label{eq:papampl}
F_{\bar pp}(z) = F^H_+(z) - F^{MO}_-(z)  +F^P(z)-F^O(z)+F^R_+(z) -F^R_-(z).
\end{equation}
For $\beta_F=\beta_{MO}=2$ and $\alpha_P(0)=\alpha_O(0)=1$ we get exactly the FMO model  of Ref. \cite{MN}.
Our aim is verify which values of   $\beta_F$ and $\beta_{MO}$, as well as which values of  $\alpha_P(0)$ and $\alpha_O(0)$ are the best for fitting all existing experimental data on $\sigma_{tot}(s)$ and $\rho(s)$.

\begin{figure}[h]
	\centering
	\includegraphics[width=1.\linewidth]{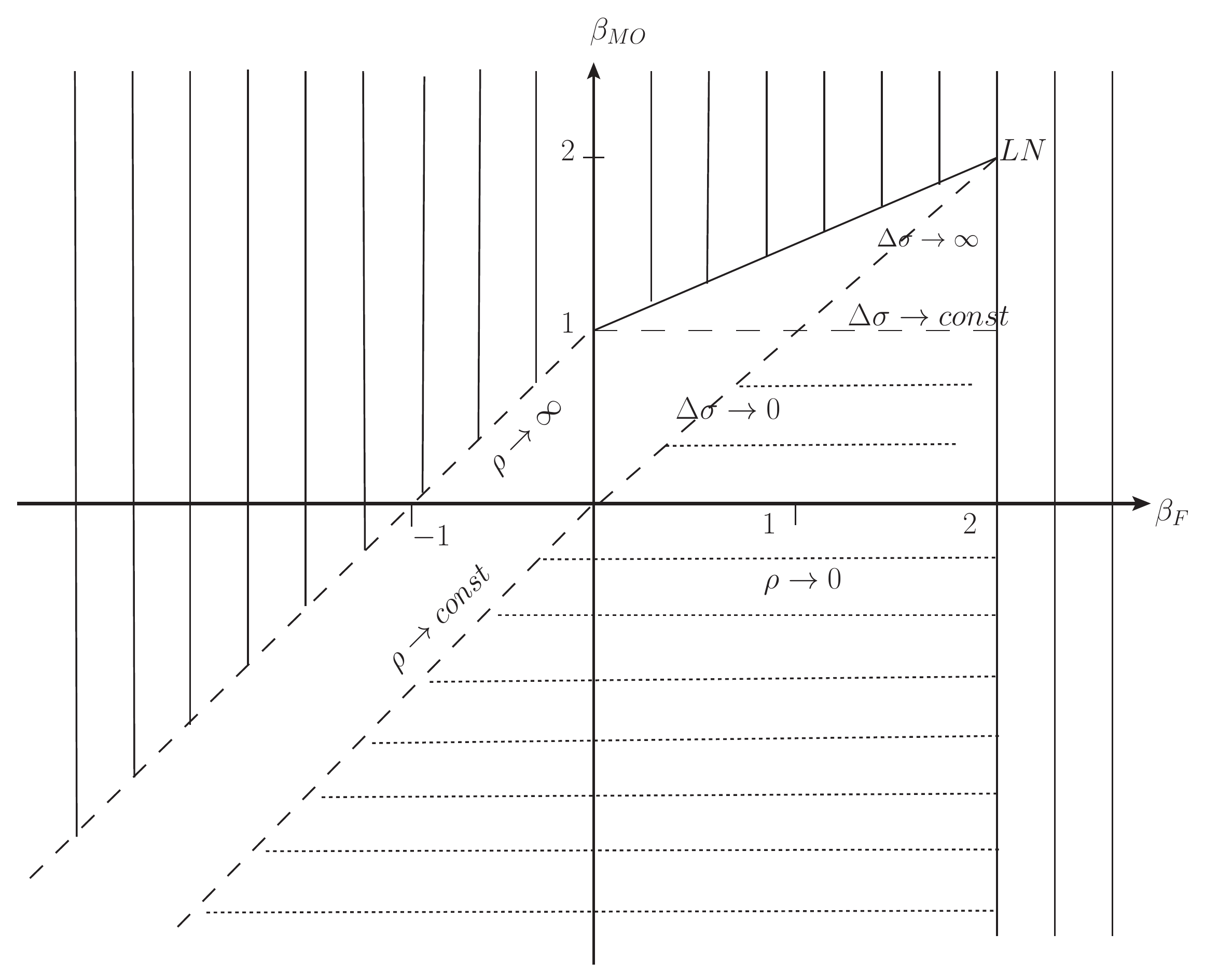}
	\caption{Cornille's plot}
	\label{fig:cornille-plot-fmo}
\end{figure}

The parameters $\beta_F$ and $\beta_{MO}$ are not arbitrary. They are constrained by analyticity, unitarity, crossing-symmetry and positivity of cross sections \cite{Cornille}. 
\begin{equation}\label{eq:beta-F,-MO}
\begin{array}{lll}
\beta_F&\leq 2,  &\quad \text{if} \quad \beta_F \geq 0, \\
\beta_{MO}&\leq \beta_F/2+1, & \quad \text{if} \quad \beta_F \geq 0, \\
\beta_{MO}&\leq \beta_F+1,  &\quad \text{if} \quad \beta_F \leq 0. 
\end{array}
\end{equation}
These three constraints can be visualized in the Cornille's plot shown in Fig. \ref{fig:cornille-plot-fmo} where we also show the different behaviors of $\Delta \sigma_=|\sigma_{tot}^{\bar pp}-\sigma_{tot}^{pp}|$ and $\rho$ in the different subregions allowed by general principles. It can be noted that the overall allowed region is strongly constrained, a fact which shows the power of general principles. It can be also be noted  that the ''conventional`` region (where $\Delta \sigma \to 0$ and $\rho \to 0$) is only a part (the hatched area of Fig. \ref{fig:cornille-plot-fmo})  of the allowed domain. The LN point in Fig. \ref{fig:cornille-plot-fmo} is precisely the point corresponding to the FMO approach ($\beta_F=\beta_{MO}=2$). 

It is also interesting to note that there are an entire class of Odderons which are different from the Maximal Odderon and might present interest on experimental level:
\begin{equation}\label{eq:bounds beta-MO} 
0\leq \beta_{MO}< 2.
\end{equation}
In particular, the case $\beta_{MO}=0$ corresponds to the Odderon Regge pole, partner of the usual Pomeron Regge pole and which leads to the behavior 
\begin{equation}
F_-(s,0)\underset{s\to \infty}{\to}\text{const}\cdot s.
\end{equation} 
It is precisely this case which was studied in 1975 in the paper in which the name ''Odderon`` was introduced for the first time \cite{JLNL}. It is a somewhat less dramatic case than the Maximal Odderon because it has no effect in $\Delta \sigma$. However, important non-asymptotic effects can be obtained at low and medium energies. 

\section{Numerical analysis}
For the analysis of the models we use the standard set of the data on $\sigma_{tot}^{pp}$, $\sigma_{tot}^{\bar pp}$  and $\rho^{pp}, \rho^{\bar pp}$ from PDG \cite{PDG}.  We also add the latest data of TOTEM at $\sqrt{s}=13$ GeV \cite{TOTEM}.
The two ATLAS data for  $\sigma_{tot}^{pp}$ \cite{ATLAS}:  $95.35\pm 1.3065$ mb at 7 TeV and  96.07$\pm0.82$ mb at 8 TeV are not included in PDG set, though they were published before PDG Review.  The data form the PDG set were taken at energies $\sqrt{s}>5 GeV$: 248 points without the two ATLAS data and 250 points with the ATLAS data. 
  
 In the Ref. \cite{MN} we excluded these two ATLAS $\sigma^{pp}$ data because they are not compatible with the TOTEM data. However in the present paper we include them in order to see their effects on the fits. We made separate fits in the first of considered models: one excluding them (M$_1$) - 248 data points  and one including them (M$'_1$) - 250 data points.
  
 We expand first the model M$_1$, defined by leaving $\beta_F$ and $\beta_{MO}$ free but keeping fixed $\alpha_P(0)=1$ and
 $\alpha_O(0)=1$. 
 
The results are quite spectacular (see the Table \ref{tab}): the values of $\beta_F$ and $\beta_{MO}$ come back to the saturation values $\beta_F=2$ and $\beta_{MO}=2$. The parameters are near those of Ref. \cite{MN}.

\begin{table}
\begin{center}
{\footnotesize 
	\begin{tabular}{|c|c|c|c|c|}
		\hline
									 & M$_0$                  &M$_1$                        &          M$'_{1}$        & M$_2$\\
		\hline                                                             
		$\beta_F$ 	          & 2.0                       &         2.0                     & 2.0                           & 2.0\\
									 & fixed                     &$\pm $0.296               &     $\pm $0.296         & fixed \\
		\hline                                                             
		$\delta_{MO}$	   & 0.                         & $0.28\cdot 10^{-7}$   & $0.108\cdot 10^{-6}$        & 0.\\
									 & fixed                     & $\pm $0.72                &      $\pm $0.72         & fixed\\
		\hline                                                             
		$\alpha_P(0)$	    & 1.0                       &         1.0                     & 1.0                           & 1.0\\
								      & fixed                    & fixed                          &   fixed                      & fixed \\
		\hline                                                             
		$\alpha_O(0)$	    & 1.0                       &         1.0                     & 1.0                        & -1.059\\
									  &  fixed                   & fixed                            & fixed                      &$\pm $0.072 \\
		\hline                                                             
		$\alpha_+(0)$	     & 0.640                  &        0.640                 & 0.674                       & 0.5294\\
									  & $\pm $0.047       &$\pm $ 0.048              &     $\pm $ 0.039         & $\pm $0.006\\
		\hline                                                             
		$\alpha_-(0)$	      & 0.272                  &        0.272                 & 0.290                        & 0.202\\
									  &  $\pm $ 0.070     & $\pm $0.069              &     $\pm $0.067          & $\pm $0.211\\
		\hline                                                             
		$H_1$(mb)	          & 0.240                  &        0.240                & 0.197                         & 0.275\\
									   &  $\pm $ 0.028     & $\pm $0.029            &     $\pm $0.027          & $\pm $0.003\\
		\hline                                                             
		$H_2$(mb)	          & -0.0837               &        -0.0829               & 0.877                         &-1.110 \\
									   &  $\pm $0.754      & $\pm $0.797               &     $\pm $0.787          & $\pm $0.034\\
		\hline                                                             
		$H_3$(mb)	          &  28.496               &        28.491                 & 22.237                       & 35.938\\
		                               & $\pm $ 5.394      &$\pm $ 5.677                &     $\pm $ 6.069         & $\pm $0.177\\
		\hline                                                             
		$O_1$(mb)	          &-0.0484                &        -0.0484                &  -0.0415                    &-0.054 \\
									   &  $\pm $ 0.011      &  $\pm $0.011              &     $\pm $0.011          & $\pm $0.005\\
		\hline                                                             
		$O_2$(mb)	          & 0.985                   &        0.985                    & 0.877                       & 1.108\\
									   &$\pm $0.212          & $\pm $0.212              &     $\pm $0.213           & $\pm $0.074\\
		\hline                                                             
		$O_3$(mb)	          &-4.758                   &        -4.758                   & -4.328                    & -5.372\\
									   & $\pm $1.030         & $\pm $1.026               &     $\pm $1.051           & $\pm $0.349\\
		\hline                                                             
		$C_P$(mb)	          & 0.                          &          0.                       & 0.                          & 0.\\
									   & fixed                      & fixed                          &        fixed                & fixed\\
		\hline                                                              
		$C_O$(mb)	          & 0.                         &          0.                       & 0.                          & 752.96\\
									    & fixed                    &fixed                            &        fixed               &$\pm $212.18 \\
		\hline                                                             
		$C_+^R$(mb)	          & 48.349                &        48.353               & 53.486                    & 52.107\\
										 &  $\pm $3.367      & $\pm $3.551              &     $\pm $3.551       & $\pm $1.293\\
		\hline                                                             
		$C_-^R$(mb)	          & 35.931                 &        35.932                & 35.469                   &42.867E \\
									     &  $\pm $4.592       & $\pm $ 4.547             &     $\pm $ 4.138    & $\pm $4.619\\
		\hline                                                             
		N of free 	                &                             &                                  &                                     & \\
		parameters	            & 10                        &        12                      & 12                                  & 12\\
		\hline                                                             
		$\chi^2_{tot}$	         & 254.70                 &        254.70               & 261.64                   & 246.97\\
		\hline                                                             
		$\chi^2/\text{dof}$	  & 1.070                    & 1.075                        &  1.100                    & 1.046\\
		\hline
	\end{tabular} 	
}
\end{center}
\caption{Here $\delta_{MO}=1+\beta_F/2-\beta_{MO}$, M$_0$ denotes the model of Ref. \cite{MN}, M$_1$ is defined in the text, The parameters of M$_1$ are get in fitting without the two ATLAS points, while those of M'$_1$ are get when they are included. M$_2$ is the variant where $\alpha_O(0)$ is left free} \label{tab}	
\end{table}

In spite of the supplementary two parameters as compared with Ref. \cite{MN}, the value of $\chi^2/\text{dof} =1.075$ is slightly bigger than 1.070, the value get in Ref. \cite{MN}. When ATLAS data are included the  $\chi^2/\text{dof}$  is, of course, worse ($\approx 1.10$)  but the best fit parameters are again not far from those of Ref. \cite{MN}.
	
The resulting curve for $pp$ total cross section is between TOTEM and ATLAS points with ATLAS points included. (parameter $H_1$ is slightly less than in the case without ATLAS points). However the most important fact is that ATLAS points have no  influence on the value of $\rho^{pp}$ at 13 TeV. It means that the Odderon effect is present even if the total $pp$ cross section is a little bit lower at LHC energies.  
	
\begin{figure}[h]
	\centering
	\includegraphics[width=0.9\linewidth]{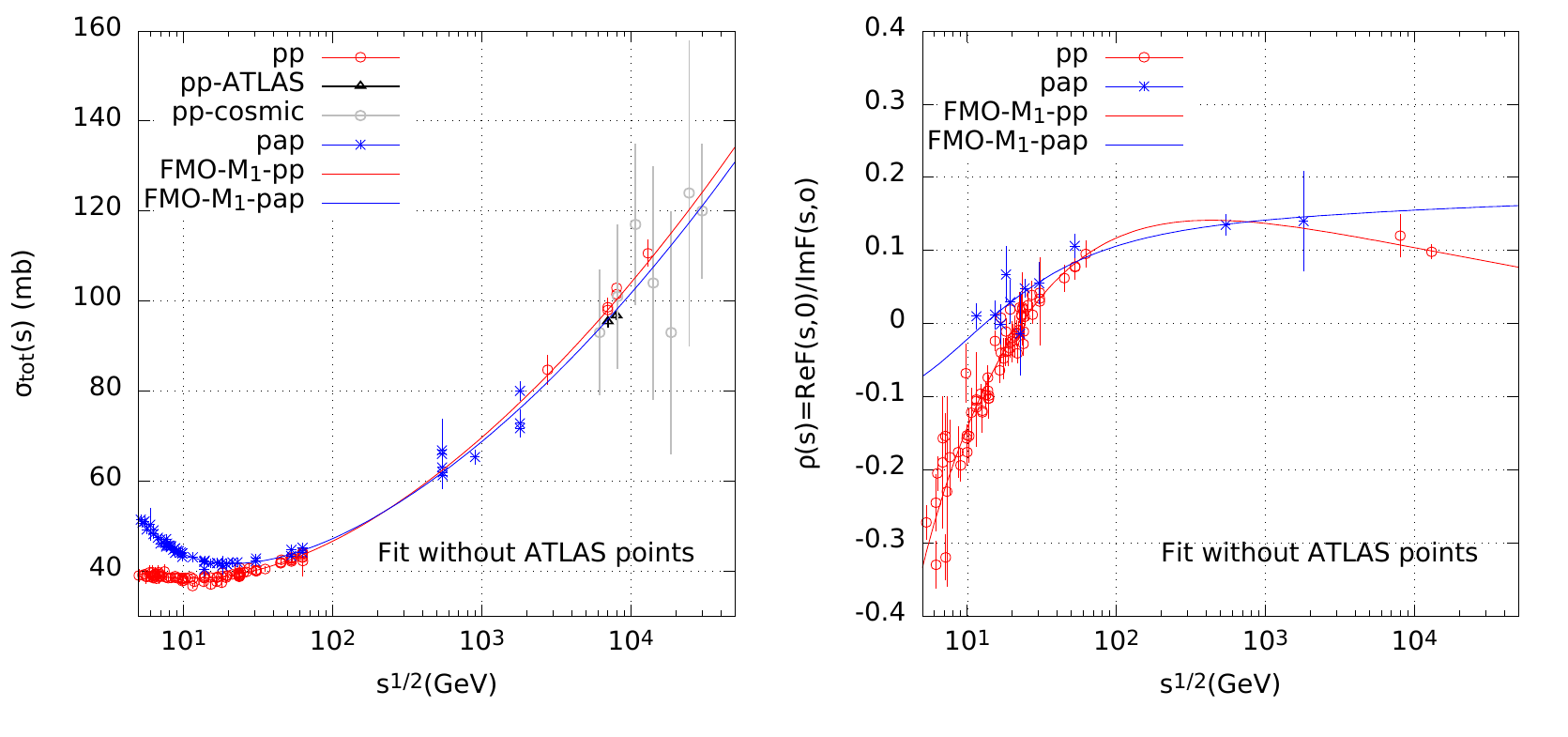}
	\caption{Model M$_1$ vs. experimental data (ATLAS data are excluded)}
	\label{fig:M1}
\end{figure}

\begin{figure}[h] 
	\centering
	\includegraphics[width=0.9\linewidth]{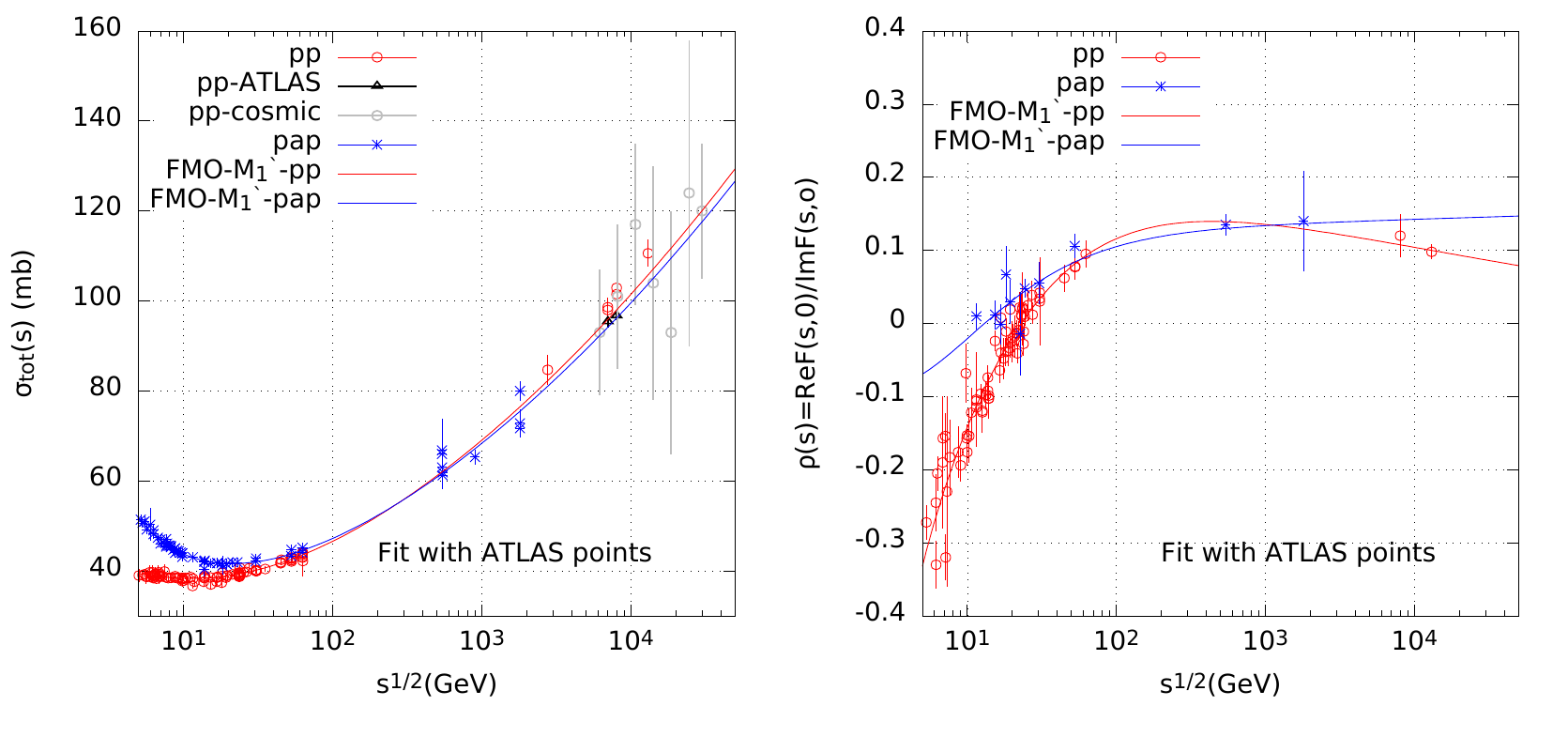}
	\caption{Model M$_1^`$ vs. experimental data (ATLAS data are included)}
	\label{fig:M1p}
\end{figure}

In Figs. \ref{fig:M1} and \ref{fig:M1p} we show the best fit Models M$_1$ and M$_1^`$  vs. experimental data. 

We made a lot of other explorations by leaving more parameters or all parameters free (with $\alpha_P(0)$ and $\alpha_O(0)$ near 1) but there are no surprises: the best fits always go back to the saturation values $\beta_F=\beta_{MO}=2$. An interesting fact is that variants with $\alpha_O(0)<1$ (but not far from $\alpha_O(0)=1$) lead to an intercept $\alpha_-(0)$ closer to the spectroscopic data, but the value of $\chi^2/\text{dof}$ is worse.

\vspace{-0.5cm}

\section{An interesting solution for Odderon Regge pole intercept $\alpha_O(0)$}

Intrigued by several calculations of the oddballs Regge trajectory in different theoretical contexts, indicating an Odderon intercept which is small and even negative (around -1) we decided to make a careful analysis of the case in which we leave $\alpha_O(0)$ completely free and not necessary near 1, while keeping $\beta_F$ and $\beta_{MO}$ at their maximal value 2. We call this variant M$_2$ (see Table \ref{tab}). 

\begin{figure}[h]
	\centering
	\includegraphics[width=1\linewidth]{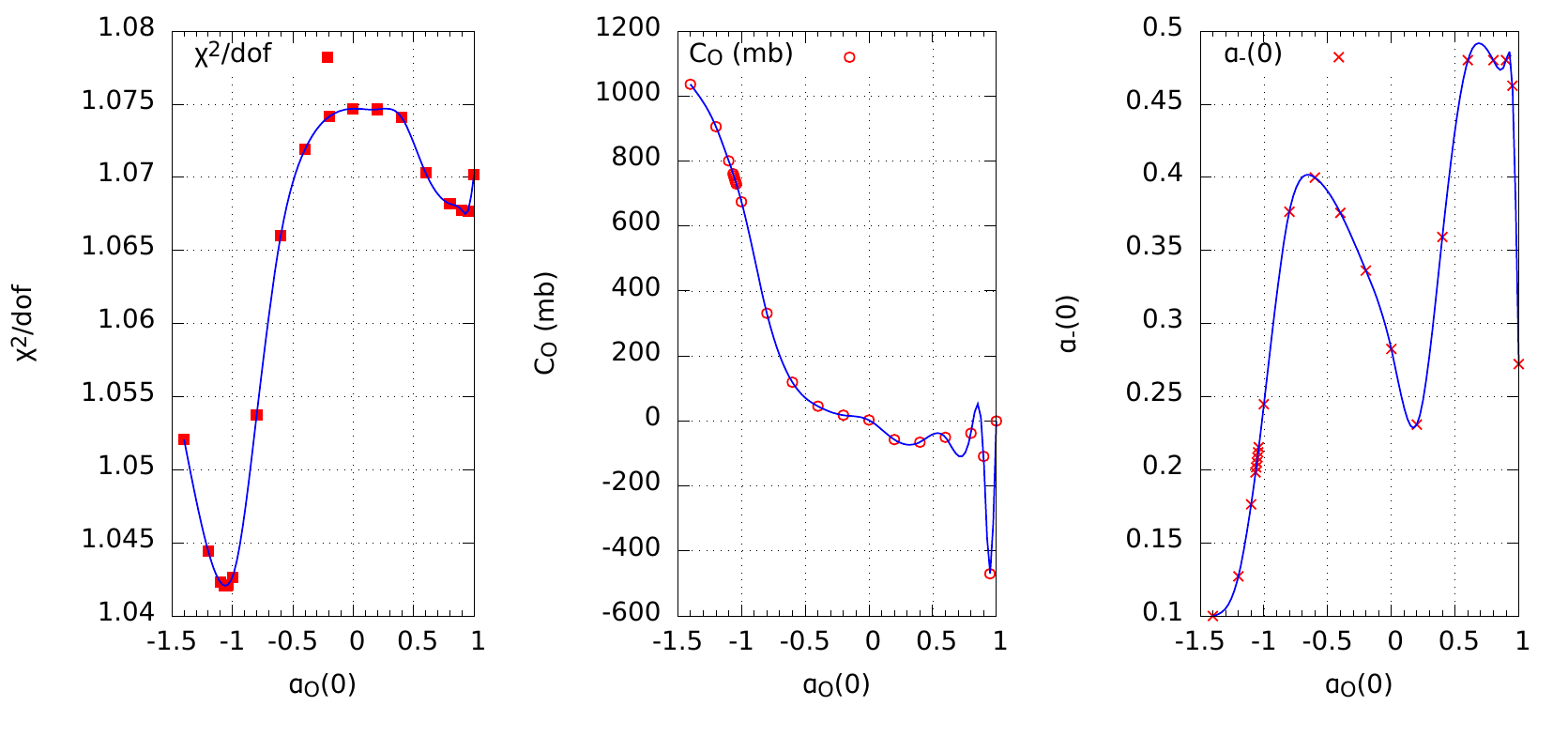}
	\caption{Dependence of $\chi^2/\text{dof}, C_O$ and $\alpha_-(0)$ on the value of Odderon Regge pole intercept}
	\label{fig:scan-Ointercept}
\end{figure}

We performed a scanning of $\chi^2$ values in terms of the intercept $\alpha_O(0)$, namely we performed fits for a set of fixed values of $\alpha_O(0)$. The results of the scanning for $\chi^2/\text{dof}$ and for parameters $C_o$ (coupling of Odderon) and $\alpha_-(0)$  (intercept of crossing-odd secondary reggeon) are shown in Fig. \ref{fig:scan-Ointercept}.

\begin{figure}[h]
	\centering
	\includegraphics[width=1.0\linewidth]{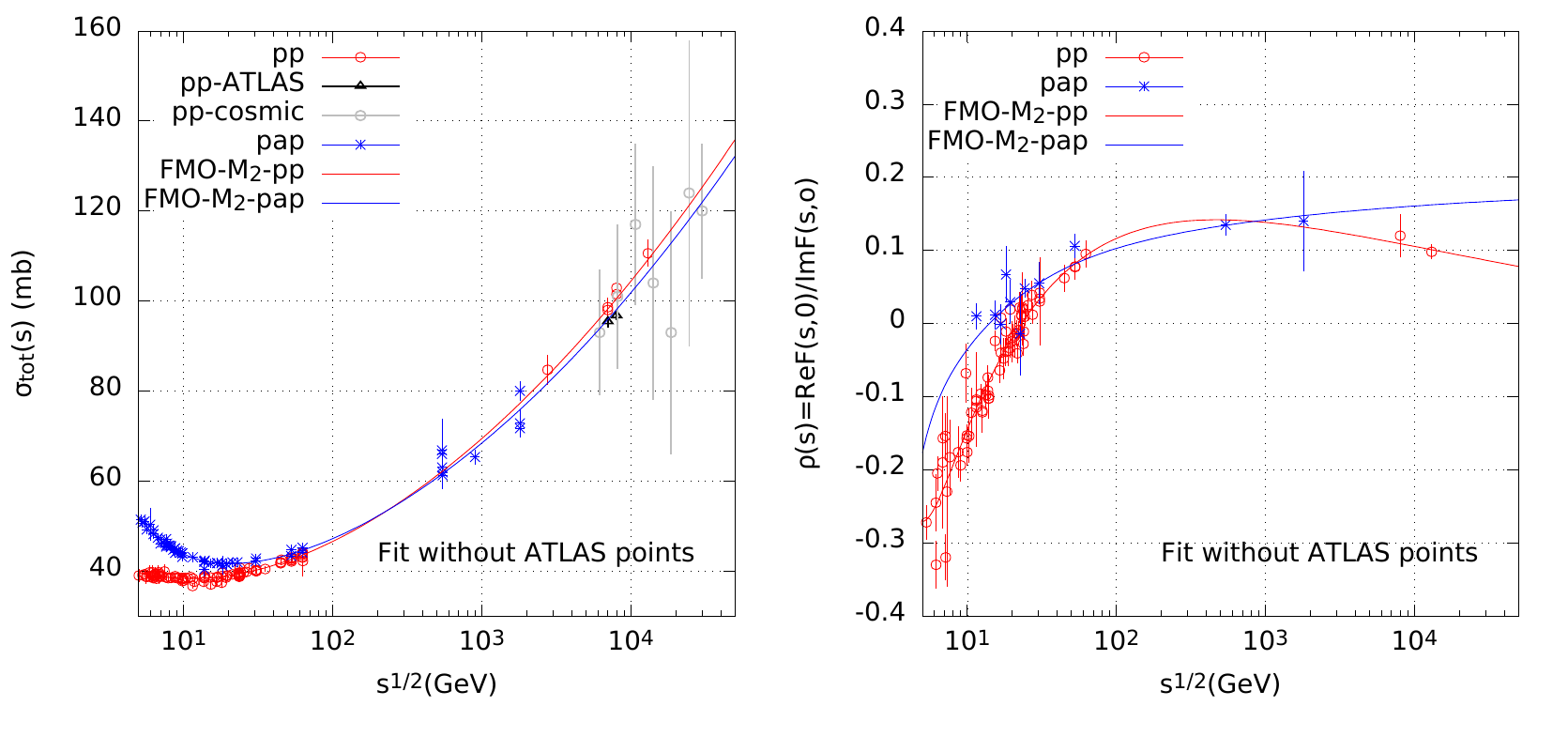}
	\caption{Model M$_2$ corresponding to low Odderon intercept vs. experimental data}
	\label{fig:M2}
\end{figure}

We were surprised to find an exceptional low value of $\chi^2/\text{dof}$ of 1.046, for a low and negative intercept $\alpha_O(0)$:
\begin{equation}\label{eq:lowOintercept}
\alpha_O(0)=-1.059\pm 0.072
\end{equation}
and
\begin{equation}\label{eq:lowOcoupling}
C_O=752.96\pm 212.18 \quad \text{mb}.
\end{equation}
The agreement of the M$_2$ variant with data is shown in Fig. \ref{fig:M2}. 

The result (\ref{eq:lowOintercept}) is in perfect agreement with the result of J. Sagedhi {\it et al.} \cite{Sagedhi} in the framework of the Odderon (oddball) spectroscopy in QCD, based on the AdS/CFT correspondence:
\begin{equation}\label{eq:SagedhiOintercept}
\alpha_O(0)=-0.9775.
\end{equation}
The agreement between these values (\ref{eq:lowOintercept}) and (\ref{eq:SagedhiOintercept})  can be hardly interpreted as a numerical coincidence. It may point to a basic structure of the 2-component Odderon: the Maximal Odderon corresponding to a dipole in imaginary part of Maximal Odderon amplitude located at $j=1$ and a simple pole located at $j=-1$. The Maximal Odderon is dominating at very high energy, while the Odderon Regge pole is important at low energy.  

 Let us mention that there are two solutions in the LLA of QCD: the Lipatov {\it et al.} solution \cite{Bartels}, located precisely at $j=1$ and corresponding to a complicated cut in the $j$-plane and the Janik-Wosiek solution \cite{Janik-Wosiek} located at $j<1$. It would be important to investigate the exact solution of the j = 1 case by going further than LLA of QCD.
 
  \section{Discussion and Conclusion}
 
 In the recent paper we showed that the LHC data strongly favor  the maximality of strong interaction and, in particular, give evidence for Maximal Odderon. In our formulation of the maximality the Maximal Odderon needs not to saturate unitarity. Saturation of asymptotic bounds  does not mean saturation of unitarity.  This  is simple to understand from physical point of view. Unitarity means the sum over all possible channels. But at infinite energies, there are a lot of non-asymptotical channels, which are not covered by Maximal Odderon. Troshin indeed verified that the Maximal Odderon does not saturate  unitarity \cite{Troshin}. Saturation of unitarity was imposed by Chew in his version of a maximality of strong interaction \cite{Chew} because Chew considered, at that moment of time, Regge physics as the final theory of strong interactions. But in the presence of singularities other than Regge poles, this assumption is no more valid.
 
  The Maximal Odderon satisfies indeed unitarity \cite{GLN}. However, a contrary claim was made recently by Khoze et al. \cite{KMR}. The authors of Ref. \cite{KMR} consider the Maximal Odderon at the limit  $A(s,b)=i$ (the regime of Black Disk Limit (BDL)) at  $s\to \infty$, in which case 
 \begin{equation}\label{eq:rho}
\text{Re}A(s.0)/\text{ImA}(s,0)\to \text{constant}\neq 0
 \end{equation}
 In the Maximal Odderon (MO) approach the condition (\ref{eq:rho}) is valid for $A(s,t=0)$ but not for the impact parameter amplitude $A(b)$  which, for MO, is $\approx O_1\Theta(R\ln s - b)$  where $|O_1|<1$. Then from the inequalities 
 \begin{equation}\label{eq:unitarity}
 1\geq |S(b)|^2=(1-\text{Im}A(b))^2+(\text{Re}A(b))^2\geq (\text{Re}A(b))^2
 \end{equation}
 we see that unitarity is violated for $\text{ReA}(b)\neq 0$  for fixed  $b$ only in the case $\text{Im}A(s,b)=2$ (unitarity limit). However, such a regime ($\text{Re}A(s.0)/\text{Im}A(s,0)\to \text{constant}\neq 0$ and $\text{Re}A(s,b)\neq 0$  for $b<R\ln s$ ) is possible only if $0<\text{Im}A(b)<2$ . Therefore MO with appropriate choice of the couplings $H_i, O_i$ does not violate the unitarity. 
  
 Concerning a pure CEP event and a solution of Finkelstein-Kajantie problem we should notice that indeed the method of unitarization considered in \cite{KMR} does not work in FMO model because $|S(b)|$ does not vanish at $s\to \infty $
 and $b< R\ln s$. It means that another procedure of unitarization of production amplitude should be applied. For example, one can sum exchanges between upper vertex and production vertex as well as exchanges between bottom and production vertexes. We should take into account that central vertex can  have zeros depending on angular and transferred momentum of all reggeons connected with the production vertex. Zeros of the vertex will lead after integration over all angular momentum $j$ to inverse powers of corresponding rapidities compensating the fast growth of cross sections. By the way similar inverse powers appear in another approach developed in \cite{MR}.  The interesting but simplified toy model with the 3-pomeron vertexes depending on angular and transferred momenta for diffraction processes was suggested long time ago in \cite{JB}.
 
 It is important to note also that the analysis of experimental data at 7 TeV \cite{AKMT} showed that BDL, $\text{Im}A(s,b=0)=1$  at this energy is exceeded. Moreover, analysis of the data at 13 TeV \cite{AKMT-2} confirms this conclusion, the effect being even more visible than at 7 TeV. It means that the survival probability $S(b)$ mentioned in \cite{KMR} does not go to $0$ at $s\to \infty $ and at any fixed $b$ because $\text{Im}A(s,b)\nrightarrow 1$ at $s\to \infty $. Consequently the method of unitarization of CEP cross section described in \cite{KMR} is in conflict with unitarity. It does not agree with unitarity provided the new experimental facts. In such a situation a new method of unitarization must be developed which takes into account the properties of $S(b)$ following from unitarity equation rather than from the some specific model which require specific properties of $S(b)$.  
 
Maximality of strong interactions stressed in the present paper is, of course, an interesting result. But we can question its general validity from a broader perspective, by taking into account the existence of models, such as the well-known Donnachie-Landshoff (DL) model, in which the Pomeron intercept is above 1 and therefore does not satisfy unitarity and maximality \cite{DL-1}. As members of the COMPETE Collaboration, we extensively studied in the past the DL model in \cite{COMP-1, COMP-2}. COMPETE elaborated a detailed method of ranking of 256 models, based on 7 statistical indicators. As it is seen from Table A2.1 of Ref. \cite{COMP-2}, the DL model has a quite modest ranking as compared with the best model involving the $\ln^2 s$ behavior of total cross-sections. Moreover, in the last publication about the subject under interest, DL presented a fit of total and differential cross-sections and of the ratio $\rho$ by their model involving a Pomeron with intercept above 1 and an Odderon described by 3-gluon exchange contribution \cite{DL-2}. One can see from Fig. 5 of Ref.\cite{DL-2} that $\rho$ at 13 TeV is about 0.12-0.13, well above the TOTEM value. Besides, the description of dips and shoulders is not sufficiently good, especially at 53 GeV and 1.8 TeV.

One last remark concerns the relation between maximality and asymptoticity. Maximality does not mean asymptoticity. We are, of course, very far from the asymptotic regime in the TeV region of energies, but the FMO approach has the virtue to describe {\it  finite-energy effects} of the asymptotic theorems. The non-asymptotic parts of the amplitudes have still important contributions at TeV energies, but the asymptotic theorems impose strong theoretical constraints on the amplitudes which are used in order to fit the experimental data. 
 
 Beyond maximality,  an interesting result when considering LHC data in conjunction with low and medium energy data is that the additional trajectory has a very low and negative intercept, in agreement with the AdS/CFT correspondence. 
 
 It is premature to extend now the FMO approach at  $t\neq 0$. The TOTEM data at 2.76 TeV are not yet published in the region of the dip, which is of great interest for the understanding of the form of the Maximal Odderon at  $t\neq 0$. The TOTEM pp data at 2.76 TeV will allow a very meaningful comparison with the D0 data at 1.96 TeV and therefore they will possibly bring further evidence for the existence of the Odderon.
 
{\bf Acknowledgment.} One of us  (E.M.) thanks the Department of Nuclear Physics and Power Engineering of the National Academy of Sciences of Ukraine for support.

\end{document}